\documentclass[12pt,draftcls,onecolumn,journal]{IEEEtran}
\usepackage{amsfonts}
\usepackage{amsfonts}
\usepackage{setspace}

\usepackage{amsmath}
\usepackage{amssymb}
\usepackage[dvips]{graphicx}
\usepackage{epsfig}

\newcommand{\tsnr}{{\text{\footnotesize{SNR}}}}
\newcommand{\tmin}{\text{min}}
\newcommand{\E}{\mathbb{E}}
\newcommand{\e}{\mathcal{E}}

\newcommand{\R}{{\sf{R}}}
\newcommand{\Pb}{\bar{P}}
\newcommand{\sta}{{\alpha^{\ast}_{\text{opt}}}}
\newcommand{\alphao}{\alpha_{\text{opt}}}
\newcommand{\ro}{r_{\text{opt}}}
\newcommand{\stro}{{{r^{\ast}_{\text{opt}}}}}
\newcommand{\dao}{\dot{\alpha}_{\text{opt}}}
\newcommand{\dro}{\dot{r}_{\text{opt}}}
\newcommand{\figsize}{0.65}
\newcommand{\rhoo}{\rho_{\text{opt}}}
\newcommand{\tsnref}{\tsnr_{\text{eff}}}
\newcommand{\tsnrefk}{\tsnr_{\text{eff},k}}
\newcommand{\tsnrefo}{\tsnr_{\text{eff,opt}}}

\newtheorem{Lem}{Theorem}

\newtheorem{Corr}{Corollary}

\begin{document}

%
\title{Energy Efficiency in the Low-SNR Regime under Queueing Constraints and Channel Uncertainty}



%

\author{\vspace{0.3cm}
\authorblockN{Deli Qiao, Mustafa Cenk Gursoy, and Senem
Velipasalar}
\thanks{The authors are with the Department of Electrical
Engineering, University of Nebraska-Lincoln, Lincoln, NE, 68588
(e-mails: dqiao726@huskers.unl.edu, gursoy@engr.unl.edu,
velipasa@engr.unl.edu).}
\thanks{This work was supported by the National Science Foundation under Grants CCF -- 0546384 (CAREER) and CNS -- 0834753.}}
%
%
%
\maketitle
\vspace{-0.8cm}
\begin{abstract}

Energy efficiency of fixed-rate transmissions is studied in the presence of queueing constraints and channel uncertainty. It is assumed
that neither the transmitter nor the
receiver has channel side information prior to transmission. The channel coefficients are
estimated at the receiver via minimum mean-square-error (MMSE) estimation with the aid of training symbols.
 It is further assumed that the system operates under statistical queueing constraints in the form of limitations on buffer
 violation probabilities. The optimal fraction of power allocated to training is
identified. Spectral efficiency--bit energy
tradeoff is analyzed in the low-power and wideband regimes by
employing the effective capacity formulation.  In particular, it is shown that the
bit energy increases without bound in the low-power regime as the average power vanishes. A similar conclusion is reached in the wideband regime if the number of noninteracting subchannels grow without bound with increasing bandwidth.
On
 the other hand, it is proven that if the number of resolvable independent paths and hence the number of noninteracting subchannels remain bounded as the available bandwidth increases, the bit energy diminishes to its minimum value in the wideband regime. For this case, expressions for the minimum bit energy and wideband slope are derived.
  Overall, energy costs of channel uncertainty and queueing constraints are identified, and the impact of multipath richness and sparsity is determined.
\\


\end{abstract}
\newpage

\setcounter{page}{1}
\begin{spacing}{1.6}
\section{Introduction}

In wireless communications, one of the main challenges in establishing reliable communications and providing quality of service guarantees is due to randomly varying channel conditions caused by mobility and changing environment. These time-varying channel conditions are often estimated in practical systems with the aid of pilot symbols albeit only imperfectly. Due to its practical significance, pilot-assisted wireless transmissions have been extensively studied in the literature. For instance, Hassibi and Hochwald in \cite{training} obtained a capacity lower bound for pilot-assisted transmission in multiple-antenna fading channels, and identified the optimal training signal type, and its power and duration. In \cite{gursoy}, the capacity and energy-efficiency of training-based transmissions are investigated and the structure of the optimal input under peak power constraints is identified. In \cite{Tong}, an overview of pilot-assisted wireless transmission techniques and their performance analyses is provided.

In many wireless communication systems, satisfying certain quality of service
(QoS) requirements is of paramount importance in providing
acceptable performance and quality. For instance, in voice over IP
(VoIP), interactive-video (e.g,. videoconferencing), and
streaming-video applications in wireless systems, latency is a key
QoS metric and should not exceed certain levels \cite{Szigeti}. Recently, effective capacity is proposed in \cite{dapeng} as a metric that can be employed to measure the performance in the presence of statistical QoS limitations. Effective
capacity formulation uses the large deviations theory and
incorporates the statistical QoS constraints by capturing the rate
of decay of the buffer occupancy probability for large queue
lengths. Hence, effective capacity can be regarded as the maximum
throughput of a system operating under limitations on the buffer
violation probability.  The analysis and application of effective capacity in various
settings has attracted much interest recently (see e.g.,
\cite{wu-downlink}--\cite{deli} and references therein). For
instance,  Tang and Zhang in \cite{tang-powerrate} considered the
effective capacity when both the receiver and transmitter know the
instantaneous channel gains, and derived the optimal power and rate
adaptation technique that maximizes the system throughput under QoS
constraints. Liu \emph{et al.}
in \cite{finite} considered fixed-rate transmission schemes and
analyzed the effective capacity and related resource requirements
for Markov wireless channel models. In this work, the
continuous-time Gilbert-Elliott channel with ON and OFF states is
adopted as the channel model while assuming the fading coefficients
as zero-mean Gaussian distributed.


In addition to the above considerations, another important concern in wireless communications is energy-efficient operation as mobile wireless systems can only be
equipped with limited energy resources.
To measure and compare the energy efficiencies of different systems
and transmission schemes, one can choose as a metric the energy
required to reliably send one bit of information.
Information-theoretic studies show that energy-per-bit requirement
is generally minimized, and hence the energy efficiency is
maximized, if the system operates at low signal-to-noise ratio
($\tsnr$) levels and hence in the low-power or wideband regimes.
Recently, Verd\'u in \cite{sergio} determined the minimum bit
energy required for reliable communication over a general class of
channels by considering the Shannon capacity formulation, and studied of the spectral efficiency--bit energy
tradeoff in the wideband regime. 
In \cite{deli} and \cite{fixed}, we incorporated the QoS limitations in the energy efficiency analysis by employing the effective capacity, rather than Shannon capacity, as the performance metric. We identified the bit energy requirements in the low-$\tsnr$ regime. In particular, in \cite{deli},
variable-rate/variable-power and variable-rate/fixed-power
transmission schemes are studied assuming the availability of perfect channel side
information (CSI) at both the transmitter and receiver or only at the
receiver. In \cite{fixed}, the performance of fixed-rate/fixed-power transmissions is investigated when the receiver has perfect CSI while the transmitter has no such knowledge.

In this paper, as a major difference from the above-cited works, we jointly consider the three major challenges in wireless systems, namely communicating under channel uncertainty, providing QoS assurances, operating energy efficiently. We assume that the
channel is not known by the transmitter and receiver prior to
transmission, and is estimated imperfectly by the receiver through
training. In our model, we incorporate statistical queueing
constraints by employing the effective capacity formulation which
provides the maximum throughput under limitations on buffer
violation probabilities for large buffer sizes. Since the
transmitter is assumed to not know the channel, fixed-rate
transmission is considered. More specifically, the contributions of the paper are the following:

\begin{enumerate}

\item We provide a framework through which energy efficiency is measured in the presence of channel uncertainty and QoS limitations in the form of queueing constraints.

\item We obtain the optimal fraction of power that needs to be allocated to training in the presence of queueing constraints.

\item We determine the bit energy levels required for operation in the low-power and wideband regimes under channel uncertainty.

\item We identify the impact of rich and sparse multipath fading on the energy efficiency when the wideband channel is imperfectly known.

\end{enumerate}


The rest of the paper is organized as follows. Section II introduces
the system model and also delineates the training and data transmission phases, and the channel estimation method. In Section III, we briefly describe the notion of
effective capacity and the spectral efficiency--bit energy tradeoff.
Energy efficiency in the low-power regime is investigated in Section
\ref{sec:lowpower}. In Section \ref{sec:wideband}, we analyze the
energy efficiency in the wideband regime. Finally, Section \ref{sec:conclusion}
provides conclusions.

\section{System Model} \label{sec:model}
\begin{figure}
\begin{center}
\includegraphics[width=\figsize\textwidth]{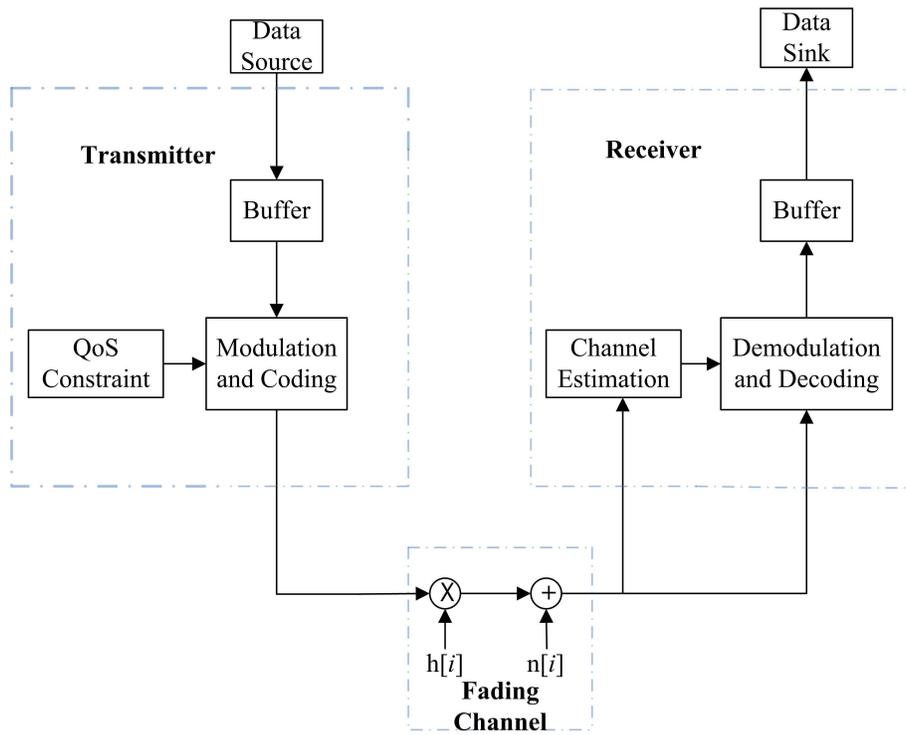}
\caption{The general system model.}\label{fig:model}
\end{center}
\end{figure}

We consider a point-to-point wireless link. Figure \ref{fig:model} illustrates the functional diagram of the system. It is assumed that the source generates
data sequences which are divided into frames of duration $T$. These
data frames are initially stored in the buffer before they are
transmitted over the wireless channel. The discrete-time channel
input-output relation in the $i^{\text{th}}$ symbol duration is
given by
\begin{gather} \label{eq:model}
y[i] = h[i] x[i] + n[i] \quad i = 1,2,\ldots.
\end{gather}
where $x[i]$ and $y[i]$ denote the complex-valued channel input and
output, respectively. We assume that the bandwidth available in the
system is $B$ and the channel input is subject to the following
average energy constraint: $\E\{|x[i]|^2\}\le \Pb / B$ for all $i$.
Since the bandwidth is $B$, symbol rate is assumed to be $B$ complex
symbols per second, indicating that the average power of the system
is limited by $\Pb$. Above in (\ref{eq:model}), $n[i]$ is a
zero-mean, circularly symmetric, complex Gaussian random variable
with variance $\E\{|n[i]|^2\} = N_0$, i.e., $n[i] \sim \mathcal{CN} (0, N_0)$. The additive Gaussian noise
samples $\{n[i]\}$ are assumed to form an independent and
identically distributed (i.i.d.) sequence. Finally, $h[i]$, which
denotes the channel fading coefficient, is assumed to be a zero-mean
Gaussian random variable with variance $E\{|h|^2\} = \gamma$. Therefore, the wireless channel is modeled as a Rayleigh fading channel. We
further assume that the fading coefficients stay constant during
 the frame duration of $T$ seconds and have independent realizations for each frame. Hence, we basically consider a block-fading channel model. Finally, we assume that neither the transmitter nor the receiver has channel side information prior to transmission. While the transmitter remains unaware of the actual realizations of the fading coefficients throughout the transmission, the receiver attempts to learn them through training.

The system operates in two phases: training phase and data
transmission phase. In the training phase, known pilot symbols are
transmitted to enable the receiver to estimate the channel
conditions, albeit imperfectly. We assume that minimum
mean-square-error (MMSE) estimation is employed at the receiver to
estimate the channel coefficient $h[i]$. Since the MMSE estimate
depends only on the training energy and not on the training duration
\cite{training} and the fading coefficients are assumed to stay constant during the frame duration of $T$ seconds, it can be easily seen that transmission of a single
pilot at every $T$ seconds is optimal. Note that in every frame
duration of $T$ seconds, we have $TB$ symbols and the overall
available energy is $\Pb T$. We now assume that each frame consists
of a pilot symbol and $TB - 1$ data symbols. The energies of the
pilot and data symbols are
\begin{equation}\label{eq:trainpower}
\e_t=\rho \Pb T, \quad\text{and}\quad \e_s=\frac{(1-\rho)\Pb
T}{TB-1},
\end{equation}
respectively, where $\rho$ is the fraction of total energy allocated
to training. Note that the data symbol energy $\e_s$ is obtained by
uniformly allocating the remaining energy among the data symbols.

In the training phase, the transmitter sends the pilot symbol $x_t = \sqrt{\e_t} = \sqrt{\rho \Pb T}$ and the receiver obtains
\begin{gather}
y_t = h \sqrt{\e_t} + n.
\end{gather}
Based on the received signal in this phase, the receiver obtains the MMSE estimate
$\hat{h} = \E\{h | y_t \}$ which can be easily seen to be a circularly symmetric, complex, Gaussian random
variable with mean zero and variance $\frac{\gamma^2 \e_t}{\gamma
\e_t + N_0}$, i.e.,
$\hat{h} \sim \mathcal{CN} \left( 0, \frac{\gamma^2 \e_t}{\gamma
\e_t + N_0} \right)$\cite{gursoy}.
Now, the channel fading coefficient $h$ can be expressed as
$h=\hat{h}+\tilde{h}$
where $\tilde{h}$ is the estimate error and $\tilde{h}\sim\mathcal
{CN}(0,\frac{\gamma N_0}{\gamma \e_t+N_0})$. Consequently, in the
data transmission phase, the channel input-output relation becomes
\begin{gather} \label{eq:impmodel}
y[i] = \hat{h}[i] x[i] + \tilde{h}[i] x[i] + n[i] \quad i =
1,2,\ldots, TB-1.
\end{gather}
Since finding the capacity of the channel in (\ref{eq:impmodel}) is
a difficult task\footnote{In \cite{gursoy}, the capacity of
training-based transmissions under input peak power constraints is
shown to be achieved by an $\tsnr$-dependent, discrete distribution
with a finite number of mass points. In such cases, no closed-form
expression for the capacity exists, and capacity values need to be
obtained through numerical computations.}, a capacity lower bound is
generally obtained by considering the estimate error $\tilde{h}$ as
another source of Gaussian noise and treating $\tilde{h}[i] x[i] +
n[i]$ as Gaussian distributed noise uncorrelated from the input \cite{training}.
Now, the new noise variance is $\E\{|\tilde{h}[i] x[i] + n[i]|^2\} =
\sigma_{\tilde{h}}^2 \e_s + N_0$ where $\sigma_{\tilde{h}}^2 =
\E\{|\tilde{h}|^2\} = \frac{\gamma N_0}{\gamma \e_t+N_0}$ is the
variance of the estimate error. Under these assumptions, a lower
bound on the instantaneous capacity is given by \cite{training},
\cite{gursoy}
\begin{align}
C_L&=\frac{TB-1}{T}\log_2\left(1+ \frac{\e_s}{\sigma_{\tilde{h}}^2
\e_s + N_0} |\hat{h}|^2\right)\\
& =\frac{TB-1}{T} \log_2\left(1+\tsnref |w|^2\right) \text{ bits/s}
\label{eq:traincap2}
\end{align}
where the effective $\tsnr$ is
\begin{equation}\label{eq:trainsnr}
\tsnref=\frac{\e_s \sigma_{\hat{h}}^2}{\sigma_{\tilde{h}}^2 \e_s +
N_0},
\end{equation}
and $\sigma^2_{\hat{h}} = \E\{|\hat{h}|^2\} = \frac{\gamma^2
\e_t}{\gamma \e_t + N_0}$ is the variance of the estimate $\hat{h}$.
Note that the expression in (\ref{eq:traincap2}) is obtained by
defining $\hat{h} = \sigma_{\hat{h}} w$ where $w$ is a standard
complex Gaussian random variable with zero mean and unit variance,
i.e., $w\sim\mathcal {CN}(0,1)$.

Since Gaussian is the worst uncorrelated noise \cite{training}, the
above-mentioned assumptions lead to a pessimistic model and   the
rate expression in (\ref{eq:traincap2}) is a lower bound to the
capacity of the true channel (\ref{eq:impmodel}). On the other hand,
$C_L$ is a good measure of the rates achieved in communication
systems that operate as if the channel estimate were perfect (i.e.,
in systems where Gaussian codebooks designed for known channels are
used, and scaled nearest neighbor decoding is employed at the
receiver) \cite{lapidoth}. Henceforth, we base our analysis on $C_L$ to understand the impact
of the imperfect channel estimate.

\begin{figure}
\begin{center}
\includegraphics[width=\figsize\textwidth]{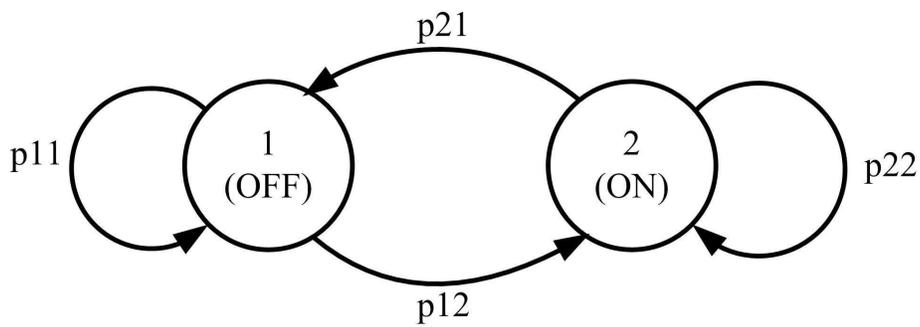}
\caption{ON-OFF state transition model.}\label{fig:00}
\end{center}
\end{figure}

Since the transmitter is unaware
of the channel conditions, it is assumed that information is
transmitted at a fixed rate of $r$ bits/s. When $r < C_L$, the
channel is considered to be in the ON state and reliable
communication is achieved at this rate. Note that under the block-fading assumption, the channel stays in the ON state for $T$ seconds and the number of  bits transmitted in this duration is $rT$. If, on the other hand, $r
\ge C_L$, we assume that outage occurs. In this case, channel is in
the OFF state during the frame duration and reliable communication at the rate of $r$ bits/s
cannot be attained. Hence, effective data rate is zero and
information has to be resent. Fig. \ref{fig:00} depicts the
two-state transmission model together with the transition
probabilities. Under the assumption of independent fading realizations in different blocks of duration $T$, it can be easily seen that the transition probabilities are
given by
\begin{align}
p_{11}&=p_{21}= P\{r \ge C_L\} =P\{|w|^2 \le \alpha\} \\
p_{22}&=p_{12}= P\{r < C_L\}= P\{|w|^2 > \alpha\}
\end{align}
where
\vspace{-.5cm}
\begin{equation}\label{eq:trainthresh}
\alpha=\frac{2^{\frac{rT}{TB-1}}-1}{\tsnref},
\end{equation}
and $|w|^2$ is an exponential random variable with mean $1$, and
hence, $P\{|w|^2 > \alpha\} = e^{-\alpha}$.

\section{Preliminaries}

\subsection{Effective Capacity}

In \cite{dapeng}, Wu and
Negi  defined the effective capacity as the maximum
constant arrival rate that a given service process can support in
order to guarantee a statistical QoS requirement specified by the
QoS exponent $\theta$ \footnote{For time-varying arrival rates,
effective capacity specifies the effective bandwidth of the arrival
process that can be supported by the channel.}. If we define $Q$ as
the stationary queue length, then $\theta$ is the decay rate of the
tail distribution of the queue length $Q$:
\begin{equation}
\lim_{q \to \infty} \frac{\log P(Q \ge q)}{q} = -\theta.
\end{equation}
Therefore, for large $q_{\max}$, we have the following approximation
for the buffer violation probability: $P(Q \ge q_{\max}) \approx
e^{-\theta q_{\max}}$. Hence, while larger $\theta$ corresponds to
more strict QoS constraints, smaller $\theta$ implies looser QoS
guarantees. Similarly, if $D$ denotes the steady-state delay
experienced in the buffer, then $P(D \ge d_{\max}) \approx
e^{-\theta \delta d_{\max}}$ for large $d_{\max}$, where $\delta$ is
determined by the arrival and service processes
\cite{Tang-crosslayer2}. Therefore, effective capacity formulation provides the maximum constant arrival rates that can be supported by the time-varying wireless channel under the queue length constraint $P(Q \ge q_{\max}) \le
e^{-\theta q_{max}}$ for large $q_{max}$ or the delay constraint $P(D \ge d_{\max}) \le
e^{-\theta \delta d_{\max}}$ for large $d_{\max}$. Since the average
arrival rate is equal to the average departure rate when the queue
is in steady-state \cite{ChangZajic}, effective capacity can also be
seen as the maximum throughput in the presence of such constraints.

The effective capacity is given by (\cite{dapeng}, \cite{chang}, \cite{cschang})
\begin{gather}
-\frac{\Lambda(-\theta)}{\theta}=-\lim_{t\rightarrow\infty}\frac{1}{\theta
t}\log_e{\mathbb{E}\{e^{-\theta S[t]}\}}
\end{gather}
where $S[t] = \sum_{i=1}^{t}R[i]$ is the time-accumulated service
process and $\{R[i], i=1,2,\ldots\}$ denote the discrete-time
stationary and ergodic stochastic service process. Note that in the
model we consider, $R[i] = rT \text{ or } 0$ depending on the
channel state being ON or OFF. In \cite{cschang}, it is shown that
for such an ON-OFF model, we have
\vspace{-.4cm}
\begin{align}
\frac{\Lambda(\theta)}{\theta}&=\frac{1}{\theta}\log_e\Big(\frac{1}{2}\Big(p_{11}+p_{22}e^{\theta
Tr}+\sqrt{(p_{11}+p_{22}e^{\theta
Tr})^2+4(p_{11}+p_{22}-1)e^{\theta Tr}} \Big)\Big).
\end{align}
Note that $p_{11}+p_{22}=1$ in our  model. Then, for a given QoS
delay constraint $\theta$, the effective capacity normalized by the
frame duration $T$ and bandwidth $B$, or equivalently spectral
efficiency in bits/s/Hz, becomes
\begin{align}
\R_E(\tsnr,\theta)&=\max_{\substack{r\geq0 \\ 0\leq
\rho\leq1}}{-\frac{1}{
TB}\frac{\Lambda(-\theta)}{\theta}} \quad \text{bits/s/Hz}\\
&=\max_{\substack{r\geq0 \\ 0\leq \rho\leq1}}{-\frac{1}{\theta
TB}\log_e\big(p_{11}+p_{22}e^{-\theta
Tr}\big)}\\
&=\max_{\substack{r\geq0 \\ 0\leq \rho\leq1}}{-\frac{1}{\theta
TB}\log_e\big(1-P(|w|^2>\alpha)(1-e^{-\theta
Tr})\big)}  \label{eq:trainopti}\\
&=-\frac{1}{\theta TB}\log_e\big(1-P(|w|^2>\alphao)(1-e^{-\theta
T\ro})\big)  \label{eq:trainopti2}.
\end{align}
Note that $\R_E$ is obtained by optimizing both the fixed
transmission rate $r$ and the fraction of power allocated to
training, $\rho$. In the optimization result
(\ref{eq:trainopti2}), $\ro$ and $\alphao$ are the optimal values
of $r$ and $\alpha$, respectively. $\ro$ can be found
by solving
\begin{equation}\label{eq:optr}
\frac{2^{\frac{Tr}{TB-1}}T\log_e2}{(TB-1)\tsnref}(1-e^{-\theta
Tr})-\theta T e^{-\theta Tr}=0
\end{equation}
where the left-hand side of (\ref{eq:optr}) is
the first derivative of the objective function in
(\ref{eq:trainopti}) with respect to $r$.

It can easily be seen that
\begin{align}
\R_E(\tsnr,0) &= \lim_{\theta \to 0} \R_E(\tsnr,\theta) =
\max_{r\ge 0} \,\,\frac{r}{B} \, P\left\{|w|^2 >
\frac{2^{\frac{rT}{TB-1}}-1}{\tsnref}\right\}.
\end{align}
Hence, as the QoS requirements relax, the maximum constant arrival
rate approaches the average transmission rate. On the other hand,
for $\theta > 0$, $\R_E < \frac{1}{B} \max_{r\ge 0} r P(|w|^2
>\alpha)$ in order to avoid violations of buffer constraints.

\subsection{Spectral Efficiency-Bit Energy Tradeoff in the Low-SNR regime}

In this paper, we focus on the energy efficiency of wireless
transmissions under the aforementioned statistical queueing
constraints. Since energy efficient operation generally requires
operation   at low-$\tsnr$ levels, our analysis throughout the paper
is carried out in the low-$\tsnr$ regime. In this regime, the
tradeoff between the normalized effective capacity (i.e, spectral
efficiency) $\R_E$ and bit energy $\frac{E_b}{N_0} =
\frac{\tsnr}{\R_E(\tsnr)}$ is a key tradeoff in understanding the
energy efficiency, and is characterized by the bit energy at zero
spectral efficiency and wideband slope provided, respectively, by
\begin{equation}\label{eq:ebresult}
\frac{E_b}{N_0}\bigg|_{\R_E = 0}
= \frac{1}{\dot{\R}_{E}(0)} \text{ and }
\mathcal{S}_0=-\frac{2(\dot{\R_E}(0))^2}{\ddot{\R_E}(0)}\log_e{2}
\end{equation}
where $\dot{\R}_E(0)$ and $\ddot{\R}_E(0)$ are the first and second
derivatives with respect to $\tsnr$, respectively, of the function
$\R_E(\tsnr)$ at zero $\tsnr$ \cite{sergio}. $\frac{E_b}{N_0}\Big|_{\R_E = 0}$ specifies the bit energy required as $\tsnr$ vanishes or equivalently as $\R_E \to 0$, while $\mathcal{S}_0$ provides the slope of the spectral efficiency curve at $\frac{E_b}{N_0}\Big|_{\R_E = 0}$. Therefore, $\frac{E_b}{N_0}\Big|_{\R_E = 0}$ and $\mathcal{S}_0$ provide a linear approximation of the spectral-efficiency vs. bit energy curve at small $\tsnr$ levels. We also note that in certain cases, the bit energy required for reliable communications diminishes with decreasing spectral efficiency, and we have $\frac{E_b}{N_0}\Big|_{\R_E =
0}=\frac{E_b}{N_0}_{\tmin}$.

\section{Energy Efficiency in the Low-Power Regime}
\label{sec:lowpower}

In this section, we analyze the spectral-efficiency vs. bit energy tradeoff in the low power regime in which the the average power of the system, $\Pb$, is small. However, before the low-power analysis, we first obtain the following result on the optimal
value of $\rho$. Note that this result is general and applies at all $\tsnr$ levels.
\begin{Lem} \label{theo:optrho}
At a given $\tsnr$ level, the optimal fraction of power $\rhoo$ that
solves (\ref{eq:trainopti}) does not depend on the QoS exponent
$\theta$ and the transmission rate $r$, and is given by
\begin{equation}\label{eq:optrho}
\rhoo=\sqrt{\eta(\eta+1)}-\eta
\end{equation}
where
\begin{gather} \label{eq:etaandsnr}
\eta=\frac{\gamma TB\tsnr+TB-1}{\gamma TB(TB-2)\tsnr} \quad \text{and} \quad
\tsnr = \frac{\Pb}{N_0B}.
\end{gather}
\end{Lem}
\emph{Proof:} From (\ref{eq:trainopti}) and the definition of
$\alpha$ in (\ref{eq:trainthresh}), we can easily see that for fixed
$r$, the only term in (\ref{eq:trainopti}) that depends on $\rho$ is
$\alpha$. Moreover, $\alpha$ has this dependency through $\tsnref$.
Therefore, $\rhoo$ that maximizes the objective function in
(\ref{eq:trainopti}) can be found by minimizing $\alpha$, or
equivalently maximizing $\tsnref$. Substituting the definitions in
(\ref{eq:trainpower}) and the expressions for $\sigma_{\hat{h}}^2$
and $\sigma_{\tilde{h}}^2$ into (\ref{eq:trainsnr}), we have
\begin{equation}\label{eq:trainsnref}
\tsnref=\frac{\e_s \sigma_{\hat{h}}^2}{\sigma_{\tilde{h}}^2 \e_s +
N_0} = \frac{\rho(1-\rho)\gamma^2T^2B^2\tsnr^2}{\rho \gamma
TB(TB-2)\tsnr+\gamma TB\tsnr+TB-1}
\end{equation}
where $\tsnr=\frac{\Pb}{N_0 B}$. Evaluating the derivative of
$\tsnref$ with respect to $\rho$ and making it equal to zero leads
to the expression in (\ref{eq:optrho}). Clearly, $\rhoo$ is
independent of $\theta$ and $r$.

Above, we have implicitly assumed that the maximization is performed
with respect to first $\rho$ and then $r$. However, the result will
not alter if the order of the maximization is changed. Note that the
objective function in (\ref{eq:trainopti})
\begin{gather}
g(\tsnref,r)= - \frac{1}{\theta
TB}\log_e\left(1-P\left(|w|^2>\frac{2^{\frac{rT}{TB-1}}-1}{\tsnref}\right)(1-e^{-\theta
Tr})\right)
\end{gather}
is a monotonically increasing function of $\tsnref$ for all $r$. It
can be easily verified that maximization does not affect the
monotonicity of $g$, and hence $\max_{r \ge 0} g(\tsnref,r)$ is
still a monotonically increasing function of $\tsnref$. Therefore,
in the outer maximization with respect to $\rho$, the choice of
$\rho$ that maximizes $\tsnref$ will also maximize $\max_{r \ge 0}
g(\tsnref,r)$, and the optimal value of $\rho$ is again given by
(\ref{eq:optrho}). \hfill$\square$

\begin{figure}
\begin{center}
\includegraphics[width=\figsize\textwidth]{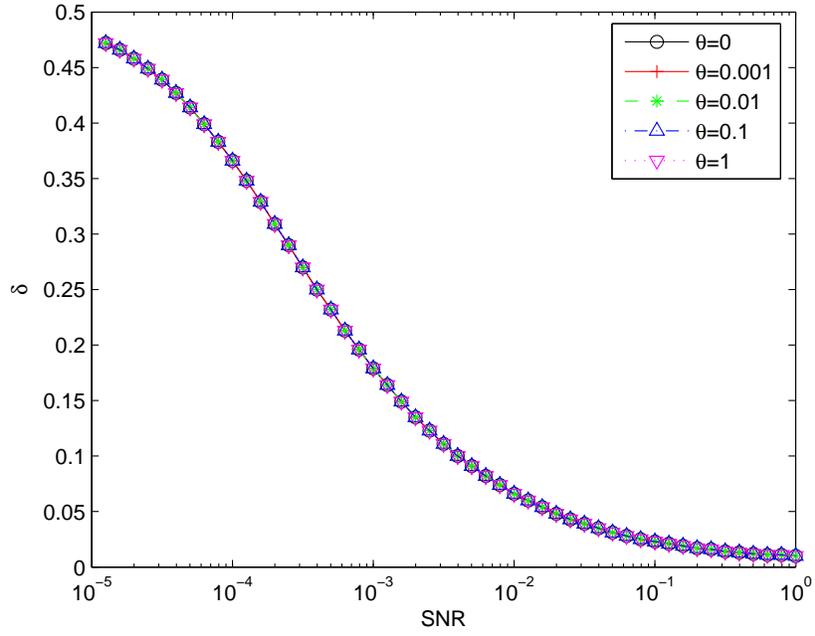}
\caption{Optimal fraction $\rhoo$ vs. $\tsnr$ in the Rayleigh
channel. $B=10^7$ Hz.}\label{fig:5}
\end{center}
\end{figure}

Fig. \ref{fig:5} plots $\rhoo$, the optimal fraction of power
allocated to training, as a function of $\tsnr$ for different values
of $\theta$ when $B=10^7$ Hz. As predicted, $\rhoo$ is the same for
all $\theta$. Note that as $\tsnr \to 0$, we have $\eta \to \infty$
and $\rhoo \to 1/2$, which is also observed in the figure. We
further observe in Fig. \ref{fig:5} that $\rhoo$ decreases with
increasing $\tsnr$. Moreover, as $\tsnr \to \infty$, we can find
that $\eta \to \frac{1}{TB-2}$ and hence $\rhoo \to
\sqrt{\frac{1}{TB-2} \left( \frac{1}{TB-2} + 1\right)} -
\frac{1}{TB-2}$. In the figure, we assume $T = 2$ ms, and therefore
$TB = 2\times 10^4$ and $\rhoo \to 0.007$.


With the optimal value of $\rho$ given in Theorem \ref{theo:optrho},
we can now express the normalized effective capacity as
\begin{align}\label{eq:Reimperf}
\R_E(\tsnr,\theta)&=\max_{r\geq0 }- \frac{1}{\theta
TB}\log_e\left(1-P\left(|w|^2>\frac{2^{\frac{rT}{TB-1}}-1}{\tsnrefo}\right)(1-e^{-\theta
Tr})\right)
\\
&=-\frac{1}{\theta
TB}\log_e\left(1-P\left(|w|^2>\frac{2^{\frac{\ro T}{TB-1}}-1}{\tsnrefo}\right)(1-e^{-\theta
T\ro})\right)\label{eq:Reimperf2}
\end{align}
where $\ro$ is the optimal value of $r$ that solves \eqref{eq:Reimperf}, and
\begin{equation}\label{eq:trainsnrefrev}
\tsnrefo=\frac{\phi(\tsnr)\tsnr^2}{\psi(\tsnr)\tsnr+TB-1},
\end{equation}
and
\begin{equation}
\phi(\tsnr)=\rhoo(1-\rhoo)\gamma^2T^2B^2,\text{ and }
\psi(\tsnr)=(1+(TB-2)\rhoo)\gamma TB.
\end{equation}%
With these notations, we obtain the following result that shows us
that operation at very low power levels is extremely energy
inefficient and should be avoided.

\begin{Lem} \label{theo:lowpower}
In the presence of channel uncertainty, the bit energy for all $\theta \ge 0$ increases
without bound as the average power $\Pb$ and hence $\tsnr$ vanishes,
i.e.,
\begin{gather}
\frac{E_b}{N_0}\bigg|_{\R_E = 0} = \lim_{\tsnr \to 0} \frac{E_b}{N_0}
= \lim_{\tsnr \to 0} \frac{\tsnr}{\R_E(\tsnr)} =
\frac{1}{\dot{\R_E}(0)} = \infty.
\end{gather}
\end{Lem}

\emph{Proof}: Recall that $|w|^2$ is an exponetial random variable with mean 1 and hence $P\{|w|^2 > \alpha\} = e^{-\alpha}$. Moreover, note that as $\tsnr \to 0$, transmission rates also approach zero and therefore we have $\ro \to 0$. Using these facts, it can be shown that the derivative of $\R_E$ in \eqref{eq:Reimperf2} with respect to $\tsnr$ at $\tsnr = 0$ is
\begin{gather}
\dot{\R}_E(0) = \lim_{\tsnr \to 0} \frac{1}{B} e^{-\alphao} \, \dot{r}_{\text{opt}}\,e^{-\theta T\ro} - \frac{1}{\theta T B} \, \dot{\alpha}_{\text{opt}} \,e^{-\alphao} (1-e^{-\theta T\ro})
\end{gather}
where $\dot{r}_{\text{opt}}$ and $\dot{\alpha}_{\text{opt}}$ are the derivatives of $\ro$ and $\alphao$, respectively, with respect to $\tsnr$, and $\alphao =\frac{2^{\frac{\ro T}{TB-1}}-1}{\tsnrefo}$. Next, we investigate how $\tsnrefo$ scales as $\tsnr$ vanishes. Note that as $\tsnr
\to 0$, $\eta \to \infty$, $\rhoo \to 1/2$, and hence $\phi(\tsnr)
\to 1/4 \gamma^2 T^2 B^2$. Then, we have
\begin{gather}
\tsnrefo = \frac{\gamma^2 T^2 B^2}{4(TB-1)} \tsnr^2 + o(\tsnr^2).
\end{gather}
Therefore, $\tsnrefo$ decreases as $\tsnr^2$ as $\tsnr$ diminishes to zero. Now, we consider the behavior of $\ro$ at low $\tsnr$s. If $\ro$ diminishes slower than $\tsnr^2$ (for instance, if $\ro$ decreases as $\tsnr^a$ where $0 < a < 2$), then it can be verified that $\alphao \to \infty$ as $\tsnr \to 0$ from which we can immediately see that $\dot{\R}_E(0) = 0$ due to exponentially decreasing term $e^{-\alphao}$. On the other hand, if $\ro$ reduces to zero faster than or as $\tsnr^2$ (e.g., as $\tsnr^a$ where $a \ge 2$), $\alphao$ approaches a finite value. However in this case, we can show that $\dot{r}_{\text{opt}} \to 0$ and $\dot{\alpha}_{\text{opt}} (1-e^{-\theta T\ro}) \to 0$ as $\tsnr \to 0$, leading again the conclusion that $\dot{\R}_E(0) = 0$.
\hfill $\square$

\emph{Remark:} Theorem \ref{theo:lowpower} shows that
$\frac{E_b}{N_0}\Big|_{\R_E = 0} = \infty$ for any $\theta \ge 0$. Hence, as noted before, operation at very low power levels is extremely energy inefficient. One reason for this behavior is that although channel estimation at very low power levels does not provide reliable estimates, the receiver regards this estimate as perfect. Hence, in the low-power regime, we have both diminishing power and deteriorating channel estimate, which affect the performance adversely. The result of Theorem \ref{theo:lowpower} also indicates that the minimum bit energy, which can be identified numerically, is achieved at a non-zero power level. In the numerical results, we will observe that both the minimum required bit energy and the other bit energy values required at a given level
of spectral efficiency increase as the QoS constraints become more stringent.

\begin{figure}
\begin{center}
\includegraphics[width=\figsize\textwidth]{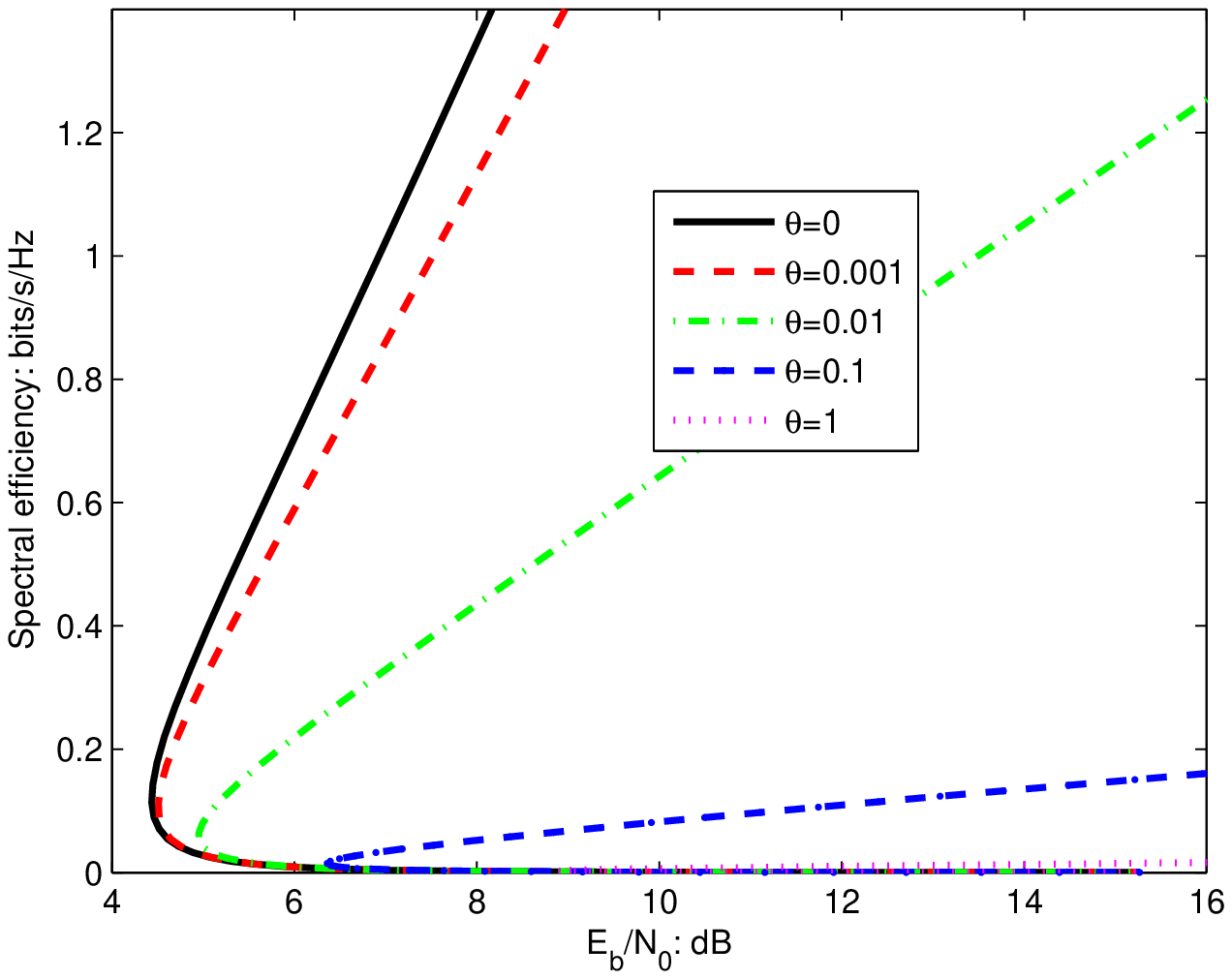}
\caption{Spectral efficiency vs. $E_b/N_0$ in the Rayleigh channel with $\E\{|h|^2\} = 1$.
$B=10^5$.}\label{fig:6}
\end{center}
\end{figure}

\begin{figure}
\begin{center}
\includegraphics[width=\figsize\textwidth]{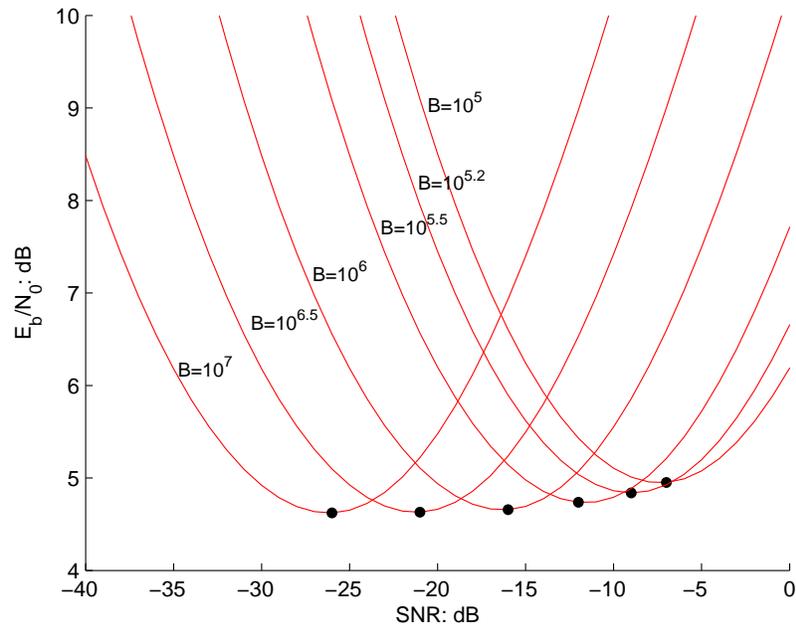}
\caption{$E_b/N_0$ vs. $\tsnr$ in the Rayleigh channel with $\E\{|h|^2\} = 1$.
$\theta$=0.01.}\label{fig:ebsnr}
\end{center}
\end{figure}

\begin{figure}
\begin{center}
\includegraphics[width=\figsize\textwidth]{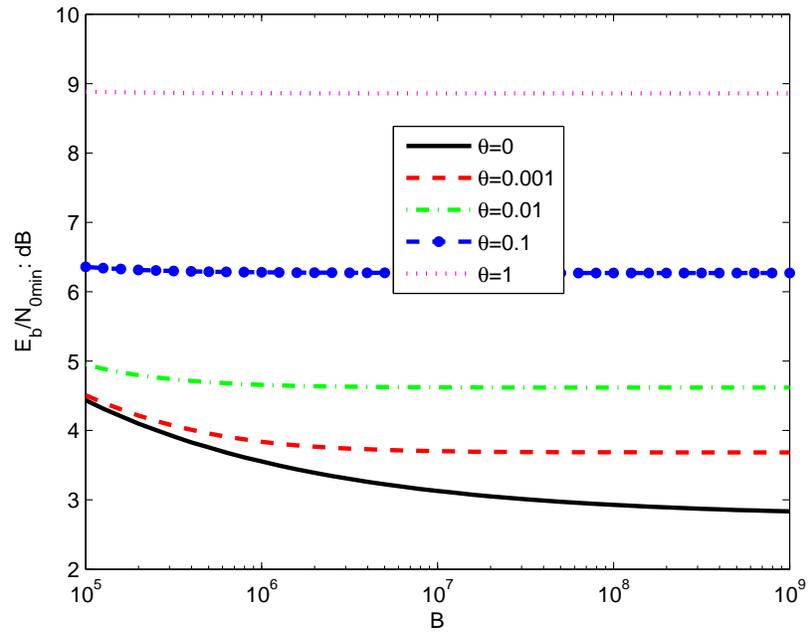}
\caption{$\frac{E_b}{N_0}_{\tmin} $vs. $B$ in the Rayleigh
channel with $\E\{|h|^2\} = 1$.}\label{fig:7}
\end{center}
\end{figure}

Fig. \ref{fig:6} plots the spectral
efficiency vs. bit energy for $\theta=\{1,0.1,0.01,0.001\}$ when
$B=10^5$ Hz in Rayleigh channel with $\E\{|h|^2\}=\gamma=1$. We
notice that as spectral efficiency $\R_E$ decreases, the bit energy $\frac{E_b}{N_0}$ initially decreases. However, as predicted by the result of
Theorem \ref{theo:lowpower}, the bit energy achieves its minimum value at a
certain nonzero spectral efficiency below which $\frac{E_b}{N_0}$ starts increasing without bound. Hence, operation below the spectral efficiency or $\tsnr$ level at which $\frac{E_b}{N_0}_{\min}$ is attained should be avoided. We also note in Fig. \ref{fig:6} that the bit energy requirements in general and the minimum bit energy in particular increases with increasing $\theta$ value, indicating the increased energy costs as the QoS limitations become more stringent.
In Fig.
\ref{fig:ebsnr}, we plot $\frac{E_b}{N_0}$ as a function of $\tsnr$ for
different bandwidth levels assuming $\theta = 0.01$. We again observe that the
minimum bit energy is attained at a nonzero $\tsnr$ value below
which $\frac{E_b}{N_0}$ requirements start increasing. Furthermore,
we see that as the bandwidth increases, the minimum bit energy tends
to decrease and is achieved at a lower $\tsnr$ level.
Finally, we plot in Fig. \ref{fig:7} the minimum bit energy as a
function of the bandwidth, $B$. We note that increasing $B$
generally decreases $\frac{E_b}{N_0}_{\min}$ value. However, there is diminishing returns as $B$ gets larger. 
Analysis in the wideband regime in the following section
will provide more insight into the impact of large bandwidth.

\section{Energy Efficiency in the Wideband Regime}
\label{sec:wideband}

In this section, we consider the wideband regime in which the
bandwidth is large. We assume that the average power $\Pb$ is kept
constant. Note that as the bandwidth $B$ increases, $\tsnr =
\frac{\Pb}{N_0B}$ approaches zero and we operate in the low-$\tsnr$
regime.

In Section \ref{sec:model}, we have described a flat fading channel model. However, flat fading assumption will not hold in the wideband regime as the bandwidth $B$ increases without bound. On the other hand, if we decompose the wideband channel
into $N$ parallel subchannels, and suppose that
each subchannel has a bandwidth that is equal to the coherence
bandwidth, $B_c$, then we can assume that independent flat-fading is experienced
in each subchannel. Note that we have $B = NB_c$. Similar to
(\ref{eq:model}), the input-output relation in the $k^{\text{th}}$
subchannel can be written as \vspace{-.4cm}
\begin{gather} \label{eq:subchannelmodel}
y_k[i] = h_k[i] x_k[i] + n_k[i] \quad i = 1,2,\ldots \quad
\text{and} \quad k = 1,2,\ldots, N.
\end{gather}
The fading coefficients $\{h_k\}_{k=1}^N$ in different subchannels
are assumed to be independent zero-mean Gaussian distributed with variances $\E\{|h_k|^2\}=\gamma_k$. The
signal-to-noise ratio in the $k^{\text{th}}$ subchannel is $\tsnr_k
= \frac{\Pb_k}{N_0 B_c}$ where $\Pb_k$ denotes the power allocated
to the $k^{\text{th}}$ subchannel and we have $\sum_{k = 1}^N \Pb_k
= \Pb$ \footnote{While not equipped with the knowledge of the instantaneous values of the fading coefficients, the transmitter is assumed to know the statistics of the fading coefficients, and possibly allocate different power levels to different subchannels with this knowledge.}. Over each subchannel, the same transmission strategy as
described in Section
 \ref{sec:model} is employed. Therefore, the transmitter, not knowing
 the fading coefficients of the subchannels, sends the data over each subchannel at the fixed rate of
 $r$. 
 Now, we can find that $C_{L,k}$ for each subchannel is given by $\frac{TB_c-1}{T} \log_2\left(1+\tsnrefk |w|^2\right) \text{ bits/s}$,
 in which
\begin{equation}\label{eq:subchanneltrainsnr}
\tsnrefk=\frac{\e_{s,k} \sigma_{\hat{h_k}}^2}{\sigma_{\tilde{h_k}}^2
\e_{s,k} + N_0}
\end{equation}
where $\e_{s,k}=\frac{(1-\rho_k) T\Pb_k}{TB_c-1}$, $\e_{t,k}=\rho_k
T\Pb_k$, $\sigma_{\tilde{h_k}}^2=\frac{\gamma_k N_0}{\gamma_k
\e_{t,k}+N_0}$ and $\sigma_{\hat{h_k}}^2=\frac{\gamma^2_k
\e_{t,k}}{\gamma_k \e_{t,k}+N_0}$.
  Similarly as before, if $r<C_{L,k}$, then transmission over the $k^{\text{th}}$ subchannel is successful.
   Otherwise, retransmission is required. Hence, we have an ON-OFF state model for each subchannel.
    On the other hand, for the transmission over $N$ subchannels, we have a state-transition model with $N+1$
    states because we have overall the following $N+1$ possible total transmission rates: $\{0, rT, 2rT, \ldots, NrT\}$.
     For instance, if all $N$ subchannels are in the OFF state simultaneously, the total rate is zero. If $j$ out of $N$ subchannels are in the ON state, then the rate is $jrT$. We note that such a decomposition strategy is also employed in \cite{fixed} where the receiver is assumed to have perfect channel information. Although similar, this strategy is also discussed here for the sake of completeness.

Now, assume that the states are enumerated in the increasing of
order of the total transmission rates supported by them. Hence, in
state $j \in \{1,\ldots,N+1\}$, the transmission rate is $(j-1)rT$.
The transition probability from state $i \in \{1,\ldots,N+1\}$ to
state $j \in \{1,\ldots,N+1\}$ is given by
\begin{align}
p_{ij} = p_{j} &= P\{(j-1) \text{ subchannels out of $N$ subchannels
are in the ON state}\}
\\
&= \sum_{\mathcal{I}_{j-1} \subset \{1, \ldots, N\}} \left(\prod_{k
\in \mathcal{I}_{j-1}} P\{|w|^2 > \alpha_k\} \prod_{k \in
\mathcal{I}^c_{j-1}} (1 - P\{|w|^2 > \alpha_k\}) \right)
\label{eq:transitionprob}
\end{align}
where $\mathcal{I}_{j-1}$ denotes a subset of the index set
$\{1,\ldots,N\}$ with $j-1$ elements. The summation in
(\ref{eq:transitionprob}) is over all such subsets. Moreover, in
(\ref{eq:transitionprob}), $\mathcal{I}^c_{j-1}$ denotes the
complement of the set $\mathcal{I}_{j-1}$, and $\alpha_k =
\frac{2^{\frac{rT}{TB_c-1}}-1}{\tsnrefk}$. Note in the above
formulation that the transition probabilities, $p_{i,j}$, do not
depend on the initial state $i$ due to the block-fading assumption.
If, in addition to being independent, the fading coefficients $h_k$
in different subchannels are identically distributed (i.e., the variances
$\{\gamma_k\}_{k=1}^N$ are the same) and also if the total power is uniformly distributed over the subchannels and the fraction of energy, $\rho_k$, allocated to training in each subchannel is the same, then $p_{i,j}$ in
(\ref{eq:transitionprob}) simplifies and becomes a binomial
probability:
\begin{align}
p_{i,j} = p_j= \left(\begin{array}{c} N \\ j-1 \end{array}\right)
\left( P\{|w|^2 > \alpha\}\right)^{j-1} \left( 1- P\{|w|^2 >
\alpha\}\right)^{N-j+1}. \label{eq:transitionprobspecial}
\end{align}
Note that with equal power allocation, we have $\Pb_k = \frac{\Pb}{N}$ and therefore $\tsnr_k =
\frac{\Pb_k}{N_0 B_c} = \frac{\Pb/N}{N_0 B/N} = \frac{\Pb}{N_0 B} =
\tsnr$ which is equal to the original $\tsnr$ used in
(\ref{eq:etaandsnr}). Since $\{\tsnrefk\}_{k=1}^N$ are also equal due to having equal $\rho_k$'s, we have the same
$\alpha=\frac{2^{\frac{rT}{TB_c-1}}-1}{\tsnref}$ for each
subchannel.

The effective capacity of this wideband channel model with N noninteracting subchannels is given by
the following result.

\begin{Lem}\label{theo:effcapwideband}
For the wideband channel with $N$ parallel noninteracting
subchannels each with bandwidth $B_c$ and independent flat fading,
the normalized effective capacity in bits/s/Hz is given by
\begin{gather}\label{eq:effectcapwideband}
\R_E(\tsnr,\theta) = \max_{\substack{r\geq0 \\ \Pb_k \ge 0 \text{ s.t. } \sum \Pb_k \le \Pb \\0\leq \rho_k\leq1 \,\,\forall k}}\left\{-\frac{1}{\theta TB} \log_e
\left( \sum_{j = 1}^{N+1} p_j \, e^{-\theta (j-1) rT} \right)
\right\}
\end{gather}
where $p_j$ is given in (\ref{eq:transitionprob}). If
$\{h_k\}_{k=1}^N$ are identically distributed Gaussian random variables with zero mean and variance $\gamma$ and the data and training energies are uniformly allocated over the subchannels, then the normalized
effective capacity expression simplifies to
\begin{gather} \label{eq:effectcapwidebandspecial}
\R_E(\tsnr,\theta) = \max_{\substack{r\geq0 \\ 0\leq \rho\leq1}}\left\{-\frac{1}{\theta
TB_c}\log_e\left(1-P\{|w|^2>\alpha\}(1-e^{-\theta
Tr})\right)\right\}.
\end{gather}
where $\alpha =\frac{2^{\frac{rT}{TB_c-1}}-1}{\tsnref}$ and $\tsnref
= \frac{\rho(1-\rho)\gamma^2T^2B_c^2\tsnr^2}{\rho \gamma
TB_c(TB_c-2)\tsnr+\gamma TB_c\tsnr+TB_c-1}$, in which
$\tsnr=\frac{\Pb}{N_0 B}=\frac{\Pb}{NN_0B_c}$.
\end{Lem}

\emph{Proof}: See \cite[Appendix A]{fixed}. 

\emph{Remark}: Theorem \ref{theo:effcapwideband} shows that if the fading coefficients in different subchannels are i.i.d. and the data and training energies are uniformly allocated over the subchannels, then the effective capacity
of a wideband channel
has an expression similar to that in (\ref{eq:trainopti}),
which provides the effective capacity of a single channel
experiencing flat fading. The only difference between
(\ref{eq:trainopti}) and (\ref{eq:effectcapwidebandspecial}) is
that $B$ is replaced in (\ref{eq:effectcapwidebandspecial}) by
$B_c$, which is the bandwidth of each subchannel.

As mentioned before, we in this section consider the wideband regime in which the
overall bandwidth of the system, $B$, is large. Additionally, we henceforth limit our analysis to the case in which the effective capacity is given by \eqref{eq:effectcapwidebandspecial} because optimization over the power allocation schemes and obtaining closed-form expressions are in general difficult tasks in the wideband regime in which the number of subchannels is potentially    high. Under these assumptions, we investigate two scenarios:
\begin{enumerate}

\item \emph{Rich multipath fading}: In this case, we assume that  the number of independent resolvable
paths increases linearly with
the bandwidth. This in turn implies that as the bandwidth $B$ increases, the number of noninteracting subchannels $N$ increases while $B_c$ stays fixed.

\item \emph{Sparse multipath fading}: In this case, we assume that  the number of independent resolvable
paths increases \emph{at most sublinearly} with
the bandwidth. This assumption implies the coherence bandwidth $B_c = \frac{B}{N}$ increases with increasing bandwidth $B$
\cite{porrat}, \cite{raghavan}. We can identify two subcases:

\begin{enumerate}
\item If the number of resolvable paths remains bounded in the wideband regime (as considered for instance in \cite{telatar}), then $N$ remains bounded while $B_c$ increases linearly with $B$.

\item If the number of resolvable paths increases but only sublinearly with $B$, then both $N$ and $B_c$ grow without bound with $B$.
\end{enumerate}
\end{enumerate}
We first consider scenario (1) where rich multipath fading is assumed. In this case, as $B$ increases, the signal-to-noise ratio $\tsnr = \frac{\Pb}{N_0 B}=\frac{\Pb}{NN_0B_c}$ approaches zero while $B_c$ stays fixed. From these facts and the similarity of the formulations in \eqref{eq:trainopti} and \eqref{eq:effectcapwidebandspecial}, we immediately conclude
that the wideband regime analysis of the rich multipath case is  the same as the low-power regime analysis conducted in Section \ref{sec:lowpower}. Therefore, as $B \to \infty$ in the rich multipath fading scenario, we have
$\frac{E_b}{N_0}\big|_{\R_E = 0} = \lim_{\tsnr \to 0} \frac{E_b}{N_0} = \infty$ for all $\theta \ge 0$. Note that we have high diversity in rich multipath fading as the number of noninteracting subchannels increase linearly with bandwidth. On the other hand, since independent fading coefficients are only imperfectly known and moreover the receiver's ability to estimate the subchannels diminishes with decreasing $\tsnr$, we have high uncertainty as well. Hence, uncertainty becomes the more dominant factor and extreme energy-inefficiency is experienced in the limit as $B \to \infty$.

Next, we
analyze the performance in the scenario of sparse multipath fading.
We note that the authors in \cite{porrat} and \cite{raghavan}, motivated by the recent measurement studies in the ultrawideband
regime, considered
sparse multipath fading channels and analyzed the performance under
channel uncertainty, employing the Shannon capacity formulation as
the performance metric.  We in this paper consider channel uncertainty and queueing constraints jointly and use the effective capacity to identify the performance. We first consider scenario (2a) where the the number of subchannels $N$ remains bounded and the degrees of freedom are limited.
The following result provides the expressions for the bit energy at
zero spectral efficiency and the wideband slope, and characterize
the spectral efficiency-bit energy tradeoff in the wideband regime when $N$ is fixed and $B_c$ grows linearly with $B$. It is shown that the bit energy required at zero spectral efficiency is indeed the minimum bit energy.
\begin{Lem} \label{theo:wideband}
For sparse multipath fading channel with bounded number of independent resolvable paths, the minimum bit energy
and wideband slope in the wideband regime are given by
\begin{gather}
\frac{E_b}{N_0}_{\tmin}=\frac{-\delta\log_e2}{\log_e\xi}
\quad \text{and} \label{eq:ebminwb}
\\
\mathcal{S}_0=\frac{\xi\log_e^2\xi \log_e2}{\theta T\sta
(1-\xi)\left(\frac{1}{T}\left(\sqrt{1+\frac{\gamma \Pb
T}{NN_0}}-1\right)+\frac{\varphi\sta}{2}\right)},\label{eq:s0wb}
\end{gather}
respectively, where $\delta=\frac{\theta T\Pb}{NN_0\log_e2}$,
$\xi=1-e^{-\sta}(1-e^{-\frac{\theta T\varphi\sta}{\log_e2}})$, and
$\varphi=\frac{\gamma \Pb}{NN_0}\left(\sqrt{1+\frac{NN_0}{\gamma \Pb
T}}-\sqrt{\frac{NN_0}{\gamma \Pb T}}\right)^2$. $\sta$ is defined as
$\sta=\lim_{\zeta\rightarrow0}\alphao$ and $\sta$ satisfies
\begin{equation} \label{eq:stacondwideband}
\sta=\frac{\log_e2}{\theta T\varphi}\log_e\left(1+\frac{\theta
T\varphi}{\log_e2}\right).
\end{equation}
\end{Lem}

\emph{Proof: } See Appendix \ref{app:wideband}.

\emph{Remark:} We note that the minimum bit energy in the sparse multipath case with bounded degrees of freedom is
achieved as $B \to \infty$ and hence as $\tsnr \to 0$. This is in
stark contrast to the results in the low-power regime and rich multipath cases in which the
bit energy requirements grow without bound as $\tsnr$
vanishes. This is due to the fact that in sparse fading with bounded number of independent resolvable paths, uncertainty does not grow without bound because the number of subchannels $N$ is kept fixed as $B \to \infty$.


\emph{Remark:} Theorem \ref{theo:wideband}, through the minimum bit
energy and wideband slope expressions, quantifies the bit energy
requirements in the wideband regime when the system is operating
subject to both statistical QoS constraints specified by $\theta$
and channel uncertainty. Note that both
$\frac{E_b}{N_0}_{\tmin}$ and $\mathcal{S}_0$ depend on
$\theta$ through $\delta$ and $\xi$.  As will be observed in the
numerical results, $\frac{E_b}{N_0}_{\min}$ and the bit energy
requirements at nonzero spectral efficiency values generally
increase with increasing $\theta$. Moreover, when compared with the
results in Section \ref{sec:lowpower}, it will be seen that sparse
multipath fading and having a bounded number of subchannels incur
energy penalty no matter there is QoS constraints or not
($\theta=0$), which is in stark contrast with previous results when
there is perfect CSI at the receiver \cite{fixed}.

\begin{figure}
\begin{center}
\includegraphics[width=\figsize\textwidth]{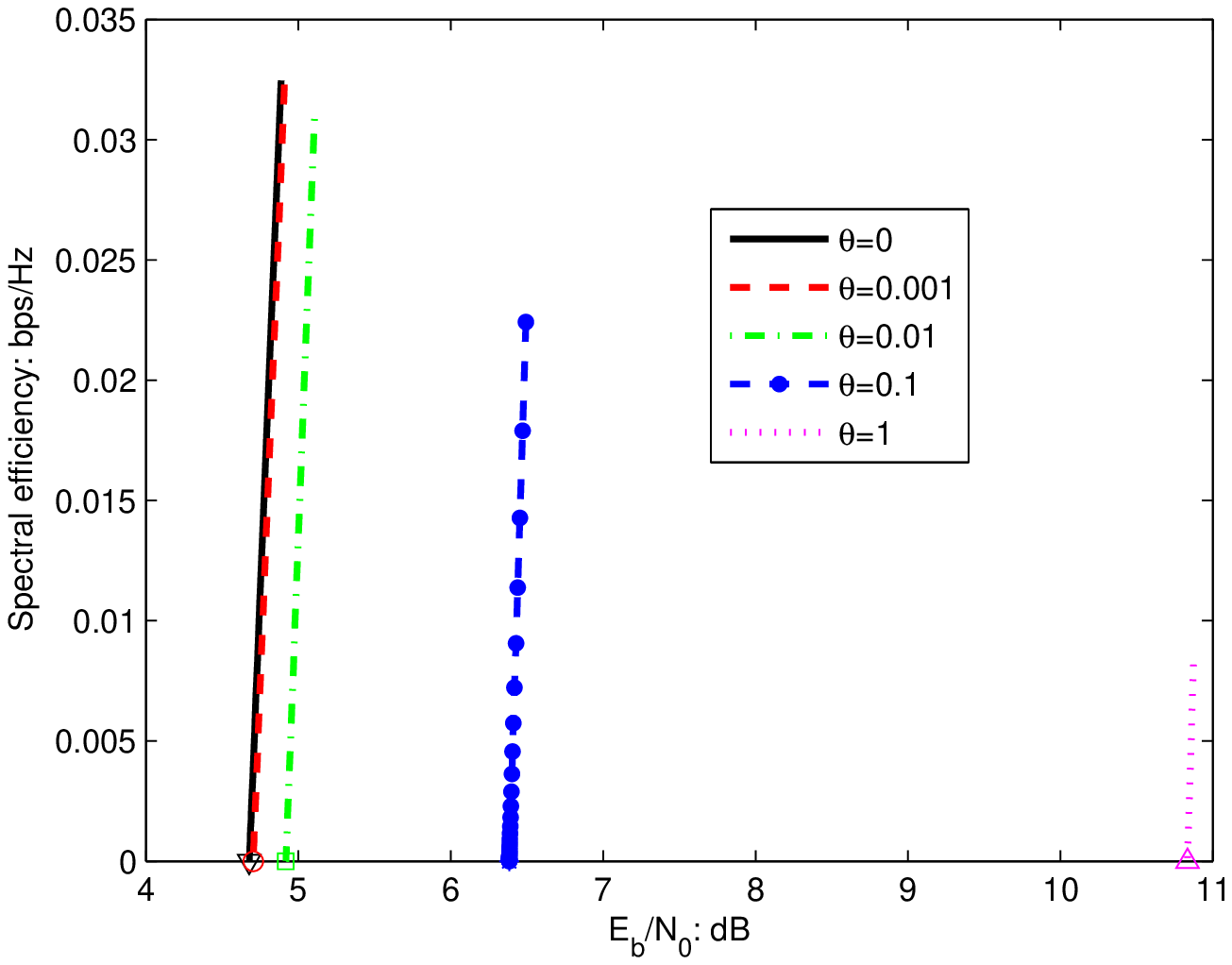}
\caption{Spectral efficiency vs. $E_b/N_0$ in the Rayleigh channel
with $E\{|h|^2\} = \gamma = 1$. $\Pb/NN_0=10^4$.}\label{fig:trainwb}
\end{center}
\end{figure}

\begin{figure}
\begin{center}
\includegraphics[width=\figsize\textwidth]{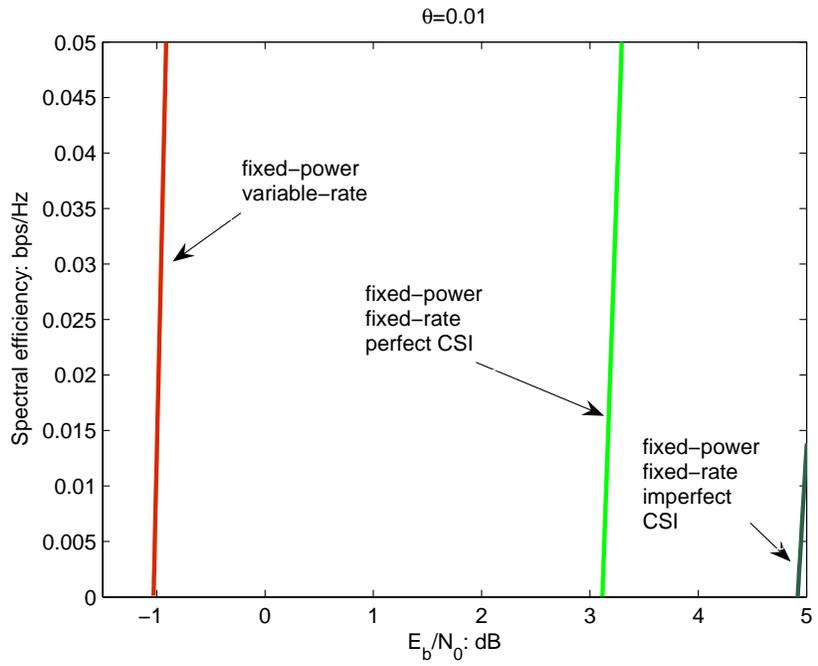}
\caption{Comparison of spectral efficiency; $\Pb/NN_0=10^4$,
$\theta=0.01$, and $E\{|h|^2\} = \gamma = 1$.}\label{fig:compair}
\end{center}
\end{figure}

After having obtained analytical expressions for the minimum bit energy and wideband slope, we now provide numerical
results. Fig. \ref{fig:trainwb} plots the spectral
efficiency--bit energy curve in the Rayleigh channel for different
$\theta$ values. In the figure, we assume that $\Pb/(NN_0)=10^4$. As
predicted, the minimum bit energies are obtained as $\tsnr$ and
hence the spectral efficiency approach zero.
$\frac{E_b}{N_0}_{\tmin}$ are computed to be equal to
$\{4.6776,4.7029,4.9177,6.3828,10.8333\}$ dB for
$\theta=\{0,0.001,0.01,0.1,1\}$, respectively. Moreover, the
wideband slopes are $\mathcal{S}_0=\{0.4720, 0.4749, 0.4978, 0.6151,
0.6061\}$ for the same set of $\theta$ values. As can also be seen
in the result of Theorem \ref{theo:wideband}, the minimum bit energy
and wideband slope in general depend on $\theta$. In Fig.
\ref{fig:trainwb}, we note that the bit energy requirements
(including the minimum bit energy) increase with increasing
$\theta$, illustrating the energy costs of stringent queueing
constraints. Finally, in this paper, we have considered
fixed-rate/fixed-power transmissions over imperfectly-known
channels. In Fig. \ref{fig:compair}, we compare the performance of
this system with those in which the channel is perfectly-known and
fixed- or variable-rate transmission is employed. The latter models
have been studied in \cite{deli} and \cite{fixed}. This figure
demonstrates the energy costs of not knowing the channel and sending
the information at fixed-rate.

We finally consider the sparse multipath fading scenario (2b) in which
the number of subchannels $N$ increases but only sublinearly with
increasing bandwidth. Note that in this case, the bit energy required as $B \to \infty$ can be obtained by
letting $N$ in the result of Theorem \ref{theo:wideband}, where $N$ is assumed to be fixed, go to
infinity.

\begin{Corr} \label{corr:sparse-diverse}
In the wideband regime, if the number of subchannels $N$ increases
sublinearly with $B$, then the bit energy required in the limit as $B \to \infty$ is
\begin{align}
\frac{E_b}{N_0}\bigg|_{\R_E=0}=\infty
\end{align}
\end{Corr}

\emph{Remark}: As $N$ increases, each subchannel is allocated
less power and operate in the low-power regime. Therefore, it is not surprising that we obtain the same bit energy result as in the low-power regime. Additionally, since the number of subchannels $N$ increases without bound, uncertainty in the channel increases as well. Hence, similarly as in rich multipath fading, extreme energy-inefficiency is experienced as $B \to \infty$.

\begin{figure}
\begin{center}
\includegraphics[width=\figsize\textwidth]{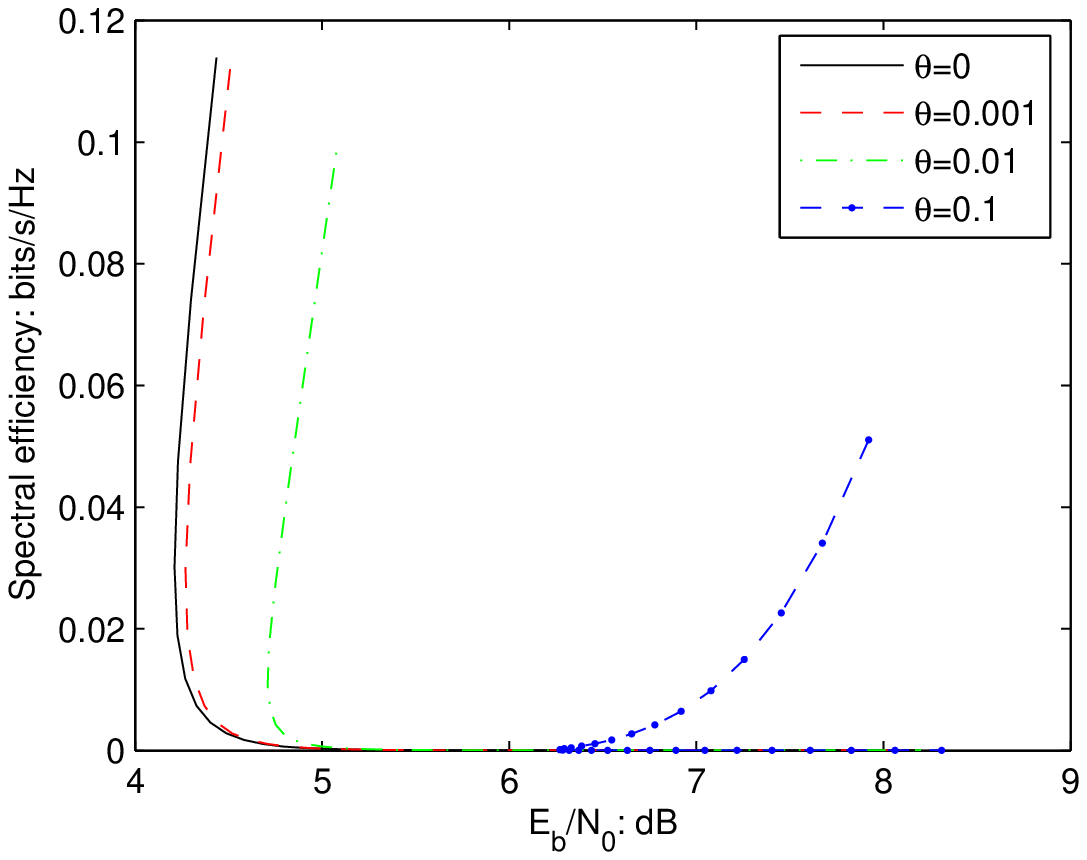}
\caption{Spectral efficiency vs. $E_b/N_0$ in the Rayleigh channel
with $E\{|h|^2\} = \gamma = 1$.
$\Pb/NN_0=10^4$.}\label{fig:imperfect}
\end{center}
\end{figure}

Fig. \ref{fig:imperfect} confirms the theoretical results. In this figure, we observe that the bit energy requirements initially decrease with decreasing spectral efficiency. However, below a certain spectral efficiency level, $\frac{E_b}{N_0}$ starts growing without bound for all $\theta \ge 0$.


\section{Conclusion} \label{sec:conclusion}

In this paper, we have analyzed the energy efficiency of fixed-rate
wireless transmissions for the communication scenario in which
queueing constraints are present and the channel coefficients are
estimated imperfectly by the receiver with the aid of training
symbols. We have considered the effective capacity as a measure of
the maximum throughput under statistical QoS constraints. We have
identified the optimal fraction of power allocated to training and
shown that this optimal fraction do not depend on the QoS exponent
$\theta$ and the transmission rate. In particular, we have
investigated the spectral efficiency--bit energy tradeoff in the
low-power and wideband regimes. 
In the low-power regime, we have shown that the bit energy increases
without bound as power diminishes. The minimum bit energy is
achieved at a certain non-zero power level below which operation
should be avoided. Although the minimum bit energy cannot be
determined in closed-form, we have observed numerically that as QoS
constraints become more stringent, the minimum bit energy increases. Similar results are obtained in the wideband regime as long as the number of subchannels increase without bound with increasing bandwidth as in rich multipath environments.
On the other hand, if the number of subchannels remains bounded as the bandwidth increases, we have shown that the bit energy required
at zero spectral efficiency (or equivalently at infinite bandwidth)
is the minimum bit energy.  We have noted that the minimum bit energy
and wideband slope in general depend on the QoS exponent $\theta$.
As the QoS constraints become more stringent and hence $\theta$ is
increased, we have observed in the numerical results that the
required minimum bit energy increases. Overall, we have quantified the
increased energy requirements in the presence of QoS constraints in the low-power and
wideband regimes, and identified the impact upon the energy
efficiency of channel uncertainty and multipath sparsity and richness.


\vspace{-0.1cm}
\appendix

\subsection{Proof of Theorem \ref{theo:wideband}} \label{app:wideband}

We first derive the following result for optimal fraction of power
on training expressed in (\ref{eq:optrho})
\begin{equation}
\rhoo=\rhoo^*+\dot{\rhoo}(0)\zeta+o(\zeta)
\end{equation}
where $\rhoo^*$ is a real value achieved as $\zeta\to0$, and
$\dot{\rhoo}(0)$ is the first derivative of $\rhoo$ evaluated at
$\zeta=0$. We have
\begin{equation}\vspace{-0.5cm}
\rhoo^*=\sqrt{\frac{NN_0}{\gamma \Pb T}\left(1+\frac{NN_0}{\gamma
\Pb T}\right)}-\frac{NN_0}{\gamma \Pb T}
\end{equation}
and
\begin{equation}
\dot{\rhoo}(0)=\frac{1}{2T}\sqrt{1+\frac{\gamma \Pb
T}{NN_0}}\left(\sqrt{1+\frac{NN_0}{\gamma \Pb
T}}-\sqrt{\frac{NN_0}{\gamma \Pb T}}\right)^2.
\end{equation}
Furthermore, $\tsnrefo$ defined below equation (\ref{eq:Reimperf})
can be simplified to
\begin{equation}\label{eq:subssnr}
\tsnrefo=\varphi\zeta+\omega\zeta^2+o(\zeta^2)
\end{equation}
where
\begin{equation}
\varphi=\frac{\rhoo^*(1-\rhoo^*)\frac{\gamma^2\Pb^2T}{(NN_0)^2}}{1+\frac{\rhoo^*\gamma
\Pb T}{NN_0}} =\frac{\gamma
\Pb}{NN_0}\left(\sqrt{1+\frac{NN_0}{\gamma \Pb
T}}-\sqrt{\frac{NN_0}{\gamma \Pb T}}\right)^2
\end{equation}
and
\begin{align}
\omega&=\frac{\frac{\gamma^2P^2T}{NN_0^2}}{1+\rhoo^*\frac{\gamma \Pb
T}{NN_0}}\bigg(\dot{\rhoo}(0)(1-2\rhoo^*)-\frac{(1-2\rhoo^*)\frac{\gamma
\Pb}{NN_0}+\dot{\rhoo}(0)\frac{\gamma \Pb
T}{NN_0}-\frac{1}{T}}{1+\frac{\rhoo^*\gamma \Pb
T}{NN_0}}\rhoo^*(1-\rhoo^*)\bigg)\nonumber\\
&=-\frac{\gamma \Pb}{NN_0T}\left(\sqrt{1+\frac{NN_0}{\gamma \Pb
T}}-\sqrt{\frac{NN_0}{\gamma \Pb
T}}\right)^2\left(\sqrt{1+\frac{\gamma \Pb T}{NN_0}}-2\right).
\end{align}
 Assume that the Taylor series expansion of $\ro$ with
respect to small $\zeta$ is
\begin{equation}\label{eq:substRb}\vspace{-0.3cm}
\ro=\stro+\dro(0)\zeta+o(\zeta)
\end{equation}
where $\stro=\lim_{\zeta\rightarrow0}\ro$ and
$\dot{r}_{\text{opt}}(0)$ is the first derivative with respect to
$\zeta$ of $\ro$  evaluated at $\zeta=0$. From
(\ref{eq:trainthresh}), we can find that
\begin{align}\vspace{-1cm}
\alphao&=\frac{2^{\frac{\ro\zeta}{1-\zeta/T}}-1}{\tsnrefo}\\
&=\frac{\stro
\log_e2+\left[\left(\frac{\stro}{T}+\dro(0)\right)\log_e2+\frac{(\stro
\log_e2)^2}{2}\right]\zeta+o(\zeta)}{\varphi+\omega\zeta+o(\zeta)}\nonumber\\
&=\frac{\stro\log_e2}{\varphi}+\bigg(\frac{\dro(0)\log_e2}{\varphi}+\frac{\stro\log_e2}{\varphi}\left(\frac{1}{T}-\frac{\omega}{\varphi}\right)+\frac{(\stro
\log_e2)^2}{2\varphi}\bigg)\zeta+o(\zeta)
\end{align}
from which we have as $\zeta\rightarrow0$ that
\begin{equation}\label{eq:stab}\vspace{-0.3cm}
\sta=\frac{\stro \log_e2}{\varphi}
\end{equation}
and that
\begin{equation}\label{eq:dotalpha}\vspace{-0.5cm}
\dao(0)=\frac{\dro(0)\log_e2}{\varphi}+\frac{\stro\log_e2}{\varphi}\left(\frac{1}{T}-\frac{\omega}{\varphi}\right)+\frac{(\stro
\log_e2)^2}{2\varphi}
\end{equation}
where $\dao(0)$ is the first derivative with respect to $\zeta$ of
$\alphao$ evaluated at $\zeta=0$. According to (\ref{eq:stab}),
$\stro=\frac{\varphi\sta}{\log_e2}$.

Combining with (\ref{eq:subssnr}) and (\ref{eq:stab}), we can obtain
from (\ref{eq:optr}) \footnote{$B$ is replaced by $B_c$ here
according to (\ref{eq:effectcapwidebandspecial}).}as $\zeta\to0$
\begin{equation}\label{eq:optcondition}\vspace{-0.2cm}
\frac{\log_e2}{\varphi}\left(1-e^{-\frac{\theta T\varphi
\sta}{\log_e2}}\right)-\theta Te^{-\theta T \stro}=0
\end{equation}
from which we get
\begin{equation}
\sta=\frac{\log_e2}{\theta T\varphi}\log_e\left(1+\frac{\theta
T\varphi}{\log_e2}\right).
\end{equation}

Since $\frac{E_b}{NN_0} =
\frac{\frac{\Pb}{NN_0}}{\frac{\R_E(\zeta)}{\zeta}}$, the result that
$\frac{E_b}{NN_0}\Big|_{\R_E = 0} = \frac{E_b}{NN_0}_{\min}$ follows
from the fact that $\R_E(\zeta)/\zeta$ monotonically decreases with
increasing $\zeta$, and hence achieves its maximum as $\zeta \to 0$.
We now have
\begin{align}
\frac{E_b}{NN_0}_{\tmin}&=\lim_{\zeta\rightarrow0}\frac{\frac{\Pb}{NN_0}
\zeta}{\R_E(\zeta)}
=\frac{-\frac{\theta
T\Pb}{NN_0}}{\log_e\big(1-P\{|w|^2\geq\sta\}(1-e^{-\theta
T\stro})\big)}\\
&=\frac{-\delta\log_e2}{\log_e\xi}=\frac{\frac{\Pb}{NN_0}}{\dot{\R}_E(0)}\label{eq:ebminb}
\end{align}
where $\dot{\R}_E(0)$ is the derivative of $\R_E$ with respect to
$\zeta$ at $\zeta = 0$, $\delta=\frac{\theta T\Pb}{NN_0\log_e2}$,
and $\xi=1-P\{|w|^2\geq\sta\}(1-e^{-\frac{\theta T
\varphi\sta}{\log_e2}})$. Obviously, (\ref{eq:ebminb}) provides
(\ref{eq:ebminwb}).

Note that the second derivative $\ddot{\R}_E(0)$, required in the
computation of the wideband slope $\mathcal{S}_0$, can be obtained
from
\begin{align}
\ddot{\R}_E(0)&=\lim_{\zeta\rightarrow0}2\frac{\R_E(\zeta)-\dot{\R}_E(0)\zeta}{\zeta^2}\nonumber\\
&=\lim_{\zeta\rightarrow0}2\frac{1}{\zeta}\Big(-\frac{1}{\theta
T}\log_e\left(1-P\{|w|^2\geq\alphao\}\left(1-e^{-\theta
T\ro}\right)\right)
+\frac{1}{\theta T}\log_e\left(1-P\{|w|^2\geq\sta\}(1-e^{-\theta
T\stro})\right)\Big)\nonumber\\
&=\lim_{\zeta\rightarrow0}-\frac{2e^{-\alphao}}{\theta
T\left(1-P\{|w|^2\geq\alphao\}\left(1-e^{-\theta
T\ro}\right)\right)}
\left(\dao(\zeta)(1-e^{-\theta T\ro})-\theta Te^{-\theta
T\ro}\dro(\zeta)\right)
\label{eq:ddotproof1}\\
&=-\frac{2e^{-\sta}}{\theta
T\left(1-P\{|w|^2\geq\sta\}\left(1-e^{-\theta
T\stro}\right)\right)}
\left(\dao(0)(1-e^{-\theta T\stro})-\theta Te^{-\theta
T\stro}\dro(0)\right)
\label{eq:ddotproof2}
\end{align}
where $\stro=\frac{\Pb\sta}{NN_0\log_e2}$. Above,
(\ref{eq:ddotproof1}) and (\ref{eq:ddotproof2}) follow by using
L'Hospital's Rule and applying Leibniz Integral Rule.

Meanwhile, substituting (\ref{eq:optcondition}) and
(\ref{eq:dotalpha}) into (\ref{eq:ddotproof2}) gives us
\begin{align}
\ddot{\R}_E(0)&=-\frac{2e^{-\sta}}{\theta
T\left(1-P\{|w|^2\geq\sta\}\left(1-e^{-\theta
T\stro}\right)\right)}\sta(1-e^{-\theta
T\stro})\left(\frac{1}{T}-\frac{\omega}{\varphi}+\frac{\varphi\sta}{2}\right)\nonumber\\
&=-\frac{2(1-\xi)\sta}{\theta
T\xi}\left(\frac{1}{T}-\frac{\omega}{\varphi}+\frac{\varphi\sta}{2}\right)\nonumber\\
&=-\frac{2(1-\xi)\sta}{\theta
T\xi}\left(\frac{1}{T}\left(\sqrt{1+\frac{\gamma \Pb
T}{NN_0}}-1\right)+\frac{\varphi\sta}{2}\right)\label{eq:ddot}
\end{align}

Combining (\ref{eq:ddot}) and (\ref{eq:ebminb}), we can prove
(\ref{eq:s0wb}). $\hfill\square$

\end{spacing}

\begin{spacing}{1.1}

\end{spacing}

\newpage

\end{document}